# Development and validation of an astronomy self-efficacy instrument for understanding and doing

Rachel Freed, David McKinnon, Michael Fitzgerald, and Christina M. Norris

*School of Education, Edith Cowan University, Joondalup Western Australia 6027, Australia*

This paper presents a new astronomy self-efficacy instrument, composed of two factors; one relating to learning astronomy content, which we call astronomy personal self-efficacy, and the other relating to the use of astronomical instrumentation, specifically the use of remote robotic telescopes for data collection. The latter is referred to as the astronomy instrumental self-efficacy factor. The instrument has been tested for reliability and construct validity. Reliability testing showed that factor 1 had a Cronbach's $\alpha$ of 0.901 and factor 2 had a Cronbach's $\alpha$ of 0.937. Construct validity was established by computing one-way analyses of variances, with the $p$ value suitably protected, using independent variables peripherally related to the constructs. These analyses demonstrate that both scales possess high construct validity. The development of this astronomy specific instrument is an important step in evaluating self-efficacy as a precursor to investigating the construct of science identity in the field of astronomy.

## I. INTRODUCTION

Improving science education is an imperative goal for current education systems in order to keep pace with technological growth and development as well as to address global issues around health and the environment. Despite decades of recognition that there are fundamental issues in science education, and numerous attempts to address these [1,2], there is still a striking lack of scientific understanding in the general population in the United States and other countries [3–5] contributing to societal norms that affect responses to global crises in such areas as climate change and viral pandemics [6]. Part of this deficiency could reasonably be attributed to a lack of interest and/or a lack of self-efficacy in science leading to low uptake of higher science courses in both high school and undergraduate courses as well as low interest in compulsory high-school science classrooms. Declining interest in science in schools has been documented numerous times over the past 20 years [3,7–10] with many studies looking for strategies to counteract this declining interest [11–14]. Studies on motivation [15–18], attitude [10,18,19], emotional impact [20], social settings [21,22], classroom environments [23–26], and other factors are important as they form the basis for the development of one's self-efficacy in a given context [27]. This further informs how best to create an education system that has real and permanent impacts, to increase personal self-efficacy for students as well as future development in the wider communities in science self-efficacy.

In a 2012 report to President Obama [28], the President's Council of Advisors on Science and Technology (PCAST) reported that the United States needed another one million science, technology, engineering, and mathematics (STEM) graduates over the next decade in order to retain its preeminence in science and technology. Specific recommendations from this report included directives to "expand the use of scientific research and engineering design courses in the first two years [of college] through a National Science Foundation (NSF) program" and to "expand opportunities for student research and design in faculty research laboratories." In order to see higher enrollments and further pursuit of science in educational and career paths, students need to both "like" science (attitudes; [3]) and "think they can do" science (self-efficacy; [29,30]). Seymour and Hewitt [31] showed that the most common reason students gave for moving out of science fields in school was a loss of interest combined with a sense of lack of relevance to their own lives.

A clear discrepancy exists between student interest in "science" versus "school science" all the way from elementary [32] to undergraduate institutions [33–35]. This has also been documented to impact undergraduate students' experience of Research Experience for Undergraduates (REUs). The REU program is defined by the Council on Undergraduate Research [36] as "A mentored investigation or creative inquiry conducted by undergraduates that seeks to make a scholarly or artistic contribution to knowledge." Studies have found that often the students expect the research experience to be like their college laboratory experience [37,38]. Early scientific research experiences, or experiences with research instrumentation, have been shown to increase students' science self-efficacy as well as their science identity [39–41] and help retain students in undergraduate science courses [42–44]. Self-efficacy is also a key mediator in identity development [39] and therefore this study developed a specific indicator for astronomy self-efficacy in both knowledge and user instrumentation.

## II. THEORETICAL FRAMEWORK

It is a well-established finding in the education literature that self-efficacy leads to course persistence [20,45,46]. It can also have a positive effect on conceptual change [47]. For instance, two of the four facets of motivation for learning in the cognitive reconstruction of knowledge model [48] include high self-efficacy within the domain of personal relevance and the social context of learning. Carlone and Johnson [49] discuss the dimensions as a juxtaposition of one's own beliefs and influence from others, as well as one's competency in their performance as drivers for science identity. Similarly, these beliefs and competencies are strongly reflected in Bandura's framework of self-efficacy.

### A. Self-efficacy

The notion of self-efficacy is constructed on the two latent factors of personal self-efficacy beliefs and outcome expectancies [29,50]. Bandura's self-efficacy framework [29] describes strong relationships between various factors that interact and influence each other. These are explained through social cognitive theory where mastery experiences, vicarious learning experiences, social persuasion experiences, and personal physiological states interact. Sawtelle *et al.* [51] found that mastery experiences are more important predictors of success in introductory physics courses for men than for women. For women, the vicarious learning experiences were greater predictors of success. The notion of the development of one's self-efficacy is context specific, and therefore would be evident under various circumstances [52,53]. In this study the context is participation in astronomy programs, which may or may not use telescopic instrumentation. Therefore, these programs provide an avenue for mastery and vicarious learning, and as they are based on teamwork would also have the potential to provide the verbal and social persuasion factors of self-efficacy. Bandura [27] stressed the importance of mastery experiences "because they provide the most authentic evidence of whether one can muster whatever it takes to succeed" (p. 80). Furthermore, research into teaching self-efficacy has also shown that emotional arousal is an important factor to allow deep engagement with the subject material [54,55]. As such, the design and tutelage of a program is paramount in influencing a positive outcome of increased self-efficacy to the future use of the learned material [56]. There is a strong mismatch between science as taught in a content-focused course and what it means to "do science" in real life. This mismatch can influence identity development in a negative manner [57–60]. As Robnett *et al.* [41] found in their longitudinal study of 251 college students, greater levels of research experience predicted higher levels of science identity, which was mediated by an increase in self-efficacy in science.

### B. Science identity

Identity can be defined as the collection of self-views that result from participation in activities and membership within a particular community [61]. Science identity involves the desire to be a "science type person" as well as socialization into the norms and discourse of science [62]. A student's science identity is an important precursor of competence beliefs [63,64]. Interest is thought by some to be a primary driver of science identity [65] and recently Colantonio *et al.* [66] described an astronomy identity framework with the four dimensions of interest, utility value, confidence, and conceptual knowledge. Applying structural equation modeling they found that interest in astronomy in middle school students has a greater effect on identity for girls than for boys and that the effect on boys was mediated through utility value. Additionally, they found that interest in astronomy, perceived utility, and identity decrease significantly with increase in grade level. Hazari and colleagues [67] found that physics identity in secondary students was exemplified through the use of the four dimensions of recognition, interest, performance, and competence. They found the development of identity is through the complex interplay of one's ability to identify self-defining characteristics and experiences, define shared experiences with others and then place these within a specific context. These contexts are strongly influenced by one's own expectations and perceptions and thus aspects of one's identity will be shaped by their own sense of self in both positive and negative ways. Again, this "sense of self" influences the construct of self-efficacy.

Identity is also key in determining retention in college STEM pathways [40]. Lopatto [68] showed through surveys given to college students and again nine months later, that participation in science research experiences increases the chance that students will persevere in a science education and career pathway. Having students use the same equipment and techniques as professional astronomers can provide authenticity for students as they are participating within a larger community of practice [69]. As students gain a sense of ownership over their data and images, they begin to see themselves authentically participating in scientific research and discourse within the larger scientific community [70], which enhances their own sense of identity as a scientist.

Astronomical images using telescopes provides vital scientific information for the development of space exploration and furthering knowledge. Research into the influence of incorporating art and imagery [71,72] to improve girls' science engagement demonstrates the importance for the use of astronomical imaging technologies to increase the future success of women in astronomy. A potential significant outcome in the study of the impact of telescope use is improving access for women and minorities to the STEM pipeline [15,21,22,73–77]. Furthermore, empirical investigations suggest that the culture of science, which tends to be exclusionary of women and their needs [78] is a potential cause of the gender gap in science performance, which first develops during the middle-school years. For secondary students, explicit discussion of the underrepresentation of women in physics positively impacted physics identity for female students. As one might expect, it had no impact on male students' physics identity [67]. Research shows the importance of having female role models in influencing STEM pathways for girls [79,80]. Furthermore, it has been found that lack of female role models and identity for women as scientists is a likely cause for the decline in motivation to pursue science [81,82]. Gonsalves [83] has reported on the lack of access to resources in physics and astronomy for women compared with men in both highly developed and less developed countries [84]. While women make up 35% of astronomy and astrophysics doctoral program enrollments, they make up only 19% of the faculty positions in the United States with no change in this percentage between the 2010 and 2014 surveys [84,85]. As reported by Barthelemy et al. [86], sexism and gender microaggressions are part of the cultural environment in physics that may discourage women from participating in physics and astronomy. It would be conceivable that the use of remote telescopes removes the observer from direct interaction with a telescope operator, thereby decreasing opportunities for stereotype threat and microaggressions and could help close the gender gap. Involving young women in astronomy research also helps to transform their identities as scientists and so provides an influence in their pursuit of STEM fields [83]. Hazari et al. [67] found that the strongest predictor of physics identity among career outcome expectation variables was a desire to pursue an intrinsically fulfilling career. They referred to this finding as a "fundamental imbalance" whereby those who come from circumstances that afford them knowledge-based motivation may opt into physics whereas those who are underrepresented often have socioeconomic motivation to pursue other pathways. Large-scale access to remote telescopes in large introductory astronomy classes or online courses has the potential to remove such negative influences for those underrepresented in astronomy. Given the desirability of developing a strong sense of science identity in such populations and the relationship between it and self-efficacy, an important first step is to develop a reliable and valid way of measuring self-efficacy.

## III. BACKGROUND

While there have been self-efficacy instruments for science teaching [20] and astronomy teaching in particular 87]], there are no self-efficacy focused instruments for "understanding" and "doing" astronomy for students. As self-efficacy is domain specific, developing a self-efficacy instrument to measure potential changes in instrumental and astronomical self-efficacy is needed. There is a general science self-efficacy factor in the Astronomy and Science Student Attitudes instrument [49] but no astronomy specific factor. Bailey et al. [88] used a five item self-efficacy instrument focused on stars and found that the greatest knowledge gains measured by the Star Properties Concept Inventory (SPCI) were related to the measured increase in self-efficacy for specific course tasks focused on learning about stars. However, there are no known self-efficacy instruments specific to general astronomy and telescope use that have been robustly validated. In general, astronomy self-efficacy has not been extensively studied.

Part of the motivation for the construction of this instrument is to probe the effect that research experiences have on participants self-efficacy for *understanding* and *doing* astronomy. With the recent expansion in interest in teacher research experiences [89,90] and undergraduate research experiences [35,37,89] it is crucial to understand the impacts of these programs and, which components are creating positive impacts. Wooten et al. [91] provide a pathway diagram, adapted from Corwin et al. [92], that presents a complex, hypothetical, flow of how, in particular, undergraduate research experiences can influence students in the short, medium and long term. This research is particularly interested in the medium and long-term outcomes, summarized in Fig. 1. As one of the key elements of the model is self-efficacy, it is necessary to have a robust validated instrument able to probe this construct.

In the context of high school, undergraduate, and teacher research experiences in astronomy, the most relevant research technology that is likely to be used is either a robotic, or remotely accessible, telescope or data that has been previously collected by such a telescope. In the context of this study, the use of this technology was of particular interest as the sample was drawn from courses that utilize these technologies. Hence the focus was on astronomy self-efficacy in general and self-efficacy in the use of the astronomical technology, robotic telescopes. The other area that would likely be used is data mining in the field commonly termed "big data," such as that used in the NASA/IPAC Teacher Archive Research Program (NITARP) [49,93,94]. As the population sample did not interact with big data, a self-efficacy scale for this technology was not attempted but this is an area that will likely benefit from a self-efficacy scale in the future.

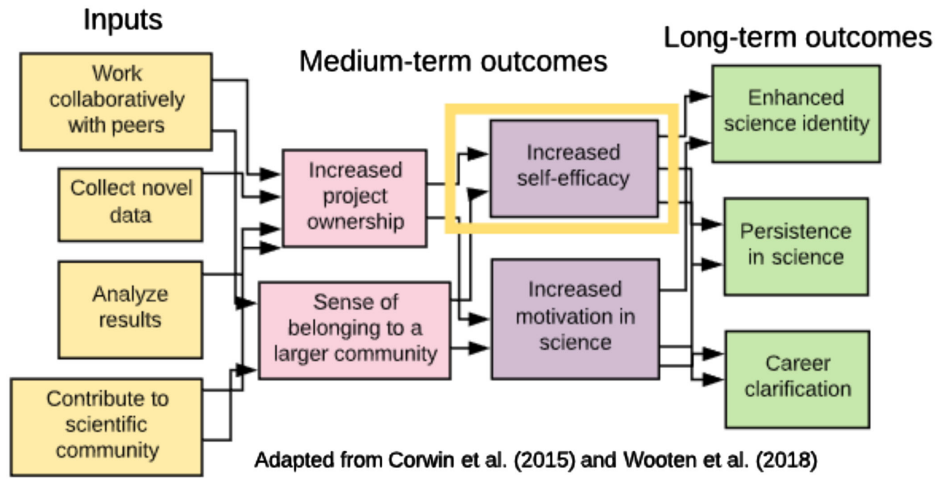

FIG. 1.   Diagram of how research experience inputs influence medium- and long-term outcomes for students. Increased self-efficacy leads to enhanced science identity, persistence in science, and increased STEM career choices.

With the general shift towards online courses, as well as public health crises mandating online learning, it could be argued that there is an even greater need for studies involving remote tools for learning in addition to remote teaching pedagogies. With the rapid growth of remote telescope technology and programs for students to use this technology [95,96], understanding the impact of these on student outcomes is important. It is a professionally shared, but as yet not robustly tested, opinion that using a research-grade telescope during a course improves students' self-efficacy and makes them feel like they "can do" astronomy [44,97,98] albeit when executed with careful focus on quality educational design [99]. Promising initial work has shown a correlation between self-efficacy and positive attitudes in robotic telescope undergraduate courses [43], however, positive shifts at the high school level have so far not been measured [100]. It is not yet clear, mainly due to a lack of explicit studies with robust instrumentation, the extent to which attitudes shift in such courses. Attitudes to astronomy, in contrast to self-efficacy, are likely to be a more difficult construct to shift, having been built up over a significant length of time over numerous experiences in an individual's life course, than more amenable constructs such as self-efficacy or content knowledge. Research on attitudes using the "Survey of Attitudes towards Astronomy" (ATA) [101] have shown a resistance to change [102,103].

## IV. STUDY CONTEXT

Throughout the United States and the rest of the world, astronomy programs are expanding significantly as evidenced by the growing participation in the Las Cumbres Observatory Global Sky Partners Program [104], as well as the increased global participation both in the Our Solar Siblings astronomy research courses [99] and the Institute for Student Astronomical Research seminars [105].

Astronomy courses are known to play the dual critical roles of being both a "gateway" science [106] and often the last science course that undergraduate students experience in their academic careers [107]. Given the focus in the literature and the importance of developing the precursor feelings of self-efficacy in the development of science identity, the specific aim of this work is to provide a new self-efficacy instrument targeted to measure the impact of astronomy programs. This allows for refinement to the designs of astronomy programs that incorporate the use of robotic telescopes. The research question at the heart of this study investigates the validity and reliability of an instrument that probes both astronomy personal self-efficacy and instrumental (the use of telescope technologies) self-efficacy as a precursor to developing science identity.

## V. METHODS

### A. Instrument development

In conducting the literature review of self-efficacy surveys in science education, we uncovered a number of STEM-related self-efficacy surveys [108–110], but none of these were specific to the use of robotic telescopes. Following Bandura's *Guide to Constructing Self-Efficacy Scales*, [30] the authors used the basis of a widely used, validated, and reliable survey known as the Science Teaching Efficacy Belief Instrument (STEBI) [23,111,112]. The STEBI-A and STEBI-B instruments involve two constructs, Personal Science Teaching Efficacy, and Science Teaching Outcomes Efficacy, which are based on a conceptual understanding of one's own capabilities and, the use of these in future teaching practices to affect students' learning. Using these broad general constructs as a guide, we developed new items that related to students' efficacy in relation to their astronomical content knowledge and to their perceived self-efficacy in being able to learn how to use the instrumentation associated with using online robotic telescopes and associated instrumentation. Thus, we developed a survey containing 27 new items theorized to probe these two aspects of "astronomy self-efficacy." Moreover, unlike the original STEBI instruments, we employed an 11-point Likert scale of 0 = Strongly Disagree to 10 = Strongly Agree as recommended by Bandura [30].

The goal was to use the constructs of one's personal astronomy conceptual understanding (the personal aspect) as a basis for the use of telescopic equipment to collect and analyze data (the outcomes). These new items are based on both basic astronomy course content and associated research on the use of robotic telescopes in education [113–115]. Data collection involved using SurveyGizmo where the survey was distributed to four undergraduate astronomy classes at the University of North Carolina, Chapel Hill in early November 2018.

### B. Data collection

Two cohorts of students (cohort A and cohort B) undertook the survey. Cohort A comprised 252 students one month into a semester-long astronomy course involving the use of robotic telescopes. Cohort B comprised 72 students who were undertaking a regular semester-long astronomy course that did not involve the use of robotic telescopes. We used cohort B as a comparison group to test the factor of the "doing" aspect of self-efficacy. Instructors allowed students to complete the surveys in class.

We examined the raw data from both cohorts to eliminate any cases with a semblance of pattern marking. In addition, we removed students from cohort B who reported having used telescopes in the past as these could invalidate the assumption that this group had not used this form of instrumentation. This left a dataset of 243 usable responses from the *lab group* and 58 from the *nonlab group*.

#### 1. Overview of statistical procedures

We employed the Statistical Package for the Social Sciences (SPSS; v26) to compute both an exploratory factor analysis (EFA) on the data obtained from cohorts A and B and also reliability analyses on the suggested interpretable factors. In addition, we computed the Cronbach's $\alpha$ for each of the groups both individually, that is, for those who had used, and those who had not used, a robotic telescope, and collectively to establish the internal consistency of responses to the items in the potential scales. Before computing the construct validity analyses of the potential scales identified by the previous two processes, we extracted a random sample of 58 cases from cohort A to match the $N$ of cohort B. Although one-way analyses of variance (ANOVA) is a robust statistical procedure for different group sizes, we wished to avoid any biases that could be otherwise controlled. This "balanced" dataset yielded an $N = 116$ cases on which we computed two

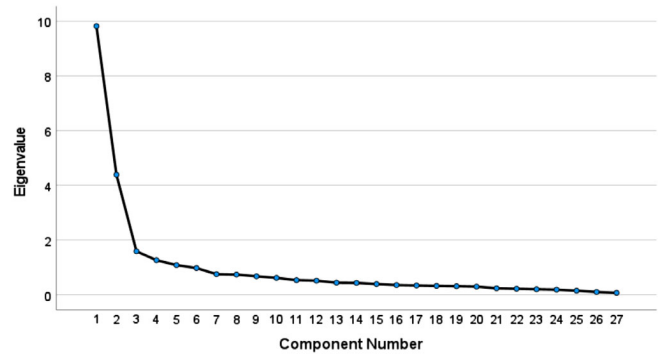

FIG. 2. Scree plot of eigenvalues.

one-way ANOVAs to establish the degree of difference for the two potential scales.

### C. Exploratory factor analysis

We computed the EFA using SPSS v26 [116] for all 324 cases from both cohort A and cohort B. The Kaiser-Meyer-Olkin (KMO) test for sampling adequacy and Bartlett's test of sphericity were computed yielding a KMO of 0.926, which indicates an adequate sample size relative to the number of test items. Bartlett's test of sphericity, based on the Chi-squared value of 5680, can reject the null hypothesis that there is no factor structure among items, $p \ll 0.0001$. We employed a principal components extraction with oblimin rotation because we hypothesized that the factors of interest would likely be correlated.

## VI. RESULTS

The initial factor solution displayed five (5) potential factors with eigenvalues greater than 1 explaining 67% of the variance. The scree plot shown in Fig. 2 clearly indicates one main factor, with an eigenvalue of 9.823 which is responsible for 36% of the variance. The abrupt change in the gradient of the scree plot at component 3 suggests that it would be useful to explore a two-, three-, and four-factor solutions [117–120].

The unconstrained factor analysis suggesting five factors produced problems in interpretation. For example, components 3, 4, and 5 showed loadings scattered across multiple components. This tends to indicate that they do not exist as separate constructs. Detailed inspection of the pattern matrix indicated that there appeared to be only two factors at play. Subsequent computations were done constraining the number of factors to 2, 3, and 4.

The computation constraining the analysis to two components yielded clearly interpretable factors. Factor 1, with an eigenvalue of 9.823 is responsible for 36.38% of the variance. Factor 2, with an eigenvalue of 4.389 is responsible for a further 16.256% of the variance. Table I shows the output for the two-factor solution. Table I indicates that both factors are negatively correlated. The correlation

TABLE I.   Pattern matrix from principal component analysis constrained to two factors.

| Pattern matrix[a] | Component 1 | Component 2 |
|---|---|---|
| Q5 After I finish this course, I feel that I can succeed if I take the next level of astronomy class | 0.835 | |
| Q6 I can explain how spectroscopy works | 0.768 | |
| Q1 I can do astronomy | 0.768 | |
| Q8 Most astronomy concepts are easy to learn | 0.742 | |
| Q9 I can explain why stars are different colors and brightnesses | 0.736 | |
| Q10 I feel that I can, with relative accuracy, visualize the universe at all different scales | 0.701 | |
| Q15 I can do the math needed in an introductory astronomy course | 0.680 | |
| Q4 I can explain how eclipses occur | 0.674 | |
| Q17 I would be able to use parallax measurements of objects within our solar system to measure the astronomical unit | 0.668 | |
| Q11 I have a good grasp of what objects exist within and around our galaxy | 0.665 | |
| Q12 The current scientific model of the origin and evolution of the universe is clear to me | 0.665 | |
| Q2 I can explain how the length of the day changes with latitude | 0.659 | |
| Q18 I can learn math well enough to be an astronomy major | 0.641 | |
| Q7 Astronomers have a solid grasp of space and time | | |
| Q25 I can show someone how to request an image from a remote telescope using an online portal | | −0.965 |
| Q24 Selecting different filters for a remote telescope observation is easy | | −0.955 |
| Q23 I am able to request telescope images through a web-based portal | | −0.940 |
| Q26 I know how to use remote telescopes | | −0.900 |
| Q21 Adjusting the brightness and contrast levels in astronomical images is straightforward | | −0.845 |
| Q27 I can learn how to use a remote telescope | | −0.738 |
| Q22 I could identify objects that are moving across the sky by examining a series of astronomical images | | −0.693 |
| Q16 Given appropriate information about standard candles (RR Lyrae, Cepheids or Type 1a Supernovae), I can calculate their distance | | −0.654 |
| Q14 I can measure angles between objects in astronomical images | 0.362 | −0.620 |
| Q13 I can distinguish between a globular cluster and galaxy in a telescope image | | −0.524 |
| Q20 Astronomers know how to use telescopes to take images of galaxies and nebula | | |
| Q19 Astronomers need to be able to do complex math | | |
| Q3 Explaining how variable stars change brightness over time is really challenging | | |

Extraction method: Principal component analysis.
Rotation method: Oblimin with Kaiser normalization.
[a]Rotation converged in seven iterations.

coefficient of these two factors is −0.264. This negative correlation is likely since students' experience of robotic telescope systems is very limited while, conversely, they have likely experienced astronomy courses during their studies at school.

## A. Factor interpretation

Factor 1 appears to be related to the construct of Astronomy Personal Self-Efficacy (APSE). Table II shows the three highest loading items together with their loading values. The items in this APSE factor all relate to identifying personally with understanding astronomy and astronomy concepts.

Factor 2 appears to be related to a construct of an instrumental component of self-efficacy. Table III shows the three most heavily loading items in relation to using a telescope through an online portal. This factor relates to how students feel about their adequacy to deal with the instrumentation aspects of using a robotic telescope in an online fashion. The items in this Instrumental Self-Efficacy (ISE) factor are more focused on one's perceived ability to interact with the remote telescopes and astronomical images leading to the doing aspect of self-efficacy.

TABLE II.   The three highest loading items contributing to factor 1.

| APSE item | EFA |
|---|---|
| After I finish this course, I feel that I can succeed if I take the next level of astronomy class | 0.835 |
| I can explain how spectroscopy works | 0.768 |
| I can do astronomy | 0.768 |

TABLE III.   The three highest loading questions contributing to factor 2.

| ISE item | EFA |
|---|---|
| I can show someone how to request an image from a remote telescope using an online portal | −0.965 |
| Selecting different filters for a remote telescope observation is easy | −0.955 |
| I am able to request telescope images through a web-based portal | −0.940 |

TABLE IV. Reliability statistics for the individual and combined cohorts.

|  | Scale | Cronbach's $\alpha$ | $F$ test | $p$ value | Tukey index |
|---|---|---|---|---|---|
| Cohort A $N = 243$ | APSE | 0.881 | $F(1, 242) = 0.008$ | $p = 0.930$ | 0.986 |
|  | ISE | 0.929 | $F(1, 242) = 1.219$ | $p = 0.270$ | 1.287 |
| Cohort B $N = 58$ | APSE | 0.814 | $F(1, 57) = 0.775$ | $p = 0.379$ | 1.221 |
|  | ISE | 0.814 | $F(1, 57) = 0.406$ | $p = 0.524$ | 1.190 |
| Cohort A + B $N = 301$ | APSE | 0.896 | $F(1, 300) = 0.012$ | $p = 0.913$ | 0.985 |
|  | ISE | 0.942 | $F(1, 300) = 0.230$ | $p = 0.631$ | 1.150 |

### B. Reliability analyses

Consistent with the principle of parsimony, we explored reducing the number of items to which participants would have to react. Rather than responding to all 27 items in the survey, it is better for respondents to respond to fewer items in a highly consistent fashion on each of the potential scales. We computed reliability analyses for the items in each of the potential scales supplied by each of the two cohorts separately since the response data were likely to differ given that one cohort had used robotic telescopes in their course and the other had not. We employed the following criteria when we assessed the reliability of the items in the two scales:

1. Cronbach's $\alpha$ had to be high for the scale ($>0.7$); and,
2. Tukey's test of additivity had to be close to 1 so that the individual item response scores (0 to 10) could be added to produce a scale.

Table IV presents the outcomes of the reliability analyses computed both separately for each cohort and for both together. The table shows Cronbach's $\alpha$, the results of the test for additivity and the Tukey index to which responses should be raised in order to achieve additivity of the individual item scores. This last statistic should be close to "1" if the responses are to be added. Otherwise, mathematical transformations are required. Previous experience as a "rule of thumb" accompanied by modeling has taught the current researchers that the range of acceptable values is $0.7 <$ Tukey index $<1.3$. Given that the EFA indicated 13 items could potentially be included in the scale, items were successively eliminated but still met the above two criteria.

Table IV shows that for the APSE potential scale for both cohorts in which eight items are employed, Cronbach's $\alpha$ is 0.896 with a nonsignificant test of additivity [$F(1, 300) = 0.012$, $p = 0.913$] and a Tukey's estimate of power to achieve additivity = 0.985. Table IV also shows the extent to which these statistics are also acceptable for the two cohorts of students separately.

Table IV shows that for the ISE potential scale comprising five items for both cohorts possess a Cronbach's $\alpha = 0.942$ and a nonsignificant test of additivity [$F(1, 300) = 0.230$, $p = 0.631$] with a Tukey's estimate of the index to achieve additivity = 1.150. Table IV also shows the extent to which these statistics are acceptable for the two cohorts of students computed separately.

These statistics for both potential scales for the entire group of respondents indicate that the raw item scores can be simply added together to produce a scale score with a high degree of internal consistency. Moreover, Table IV also shows that the potential scales for the separate cohorts of students A and B who had, respectively, used and not used robotic telescopes, also behave well in terms of reliability and additivity. We used SPSS to compute scale scores for these two factors by simply adding the relevant individual item scores together to produce a total with the range of scores for APSE being 0 to 80 and the range for ISE being 0 to 50.

### C. Construct validity analyses

We tested these two scales to see if they measure what we hypothesized them to measure. We explored the construct validity of the scales using items as independent variables that did not load onto either of the factors. To do this, several one-way ANOVA were computed. Given that the researchers computed four separate univariate computations, the $p$ value was modified using a full Bonferroni protection, that is, the $p$ value of 0.05 was divided by 4 and had to be less than 0.0125 to be recognized as potentially significant. When more than two groups were present in the IV, *post hoc* multiple comparisons with Student-Newman-Keuls protection [121] of the $p$ value were computed to examine further the extent to which the hypotheses appeared to be consistent with expectation. We give a full explanation of this multiple-comparisons procedure in the context of interpreting the output in the next section.

#### 1. Testing the construct validity of the Astronomical Personal Self-Efficacy scale

We hypothesized that those who have a higher sense of APSE would be more likely to strongly agree or agree with the statement that "*I can explain how eclipses occur*" in comparison with those who did not possess a sense of high astronomical personal self-efficacy. The null hypothesis is thus that there is no difference in the mean scale scores of those who responded differently to the independent variable. We computed a one-way ANOVA with *post hoc* multiple comparisons to test this hypothesis.

TABLE V.　Multiple comparisons using SNK for homogeneous subgroups of APSE based on the item indicated. Note that means for groups in homogeneous subsets are displayed.

| | | APSE | | | | |
|---|---|---|---|---|---|---|
| | | Subset for $\alpha = 0.05$ | | | | |
| I can explain how eclipses occur | N | 1 | 2 | 3 | 4 | 5 |
| 0 Very strongly disagree | 3 | 7.333 | | | | |
| 1 | 4 | 11.500 | | | | |
| 2 | 5 | 20.800 | 20.800 | | | |
| 3 | 10 | 22.400 | 22.400 | | | |
| 4 | 7 | | 32.429 | 32.429 | | |
| 5 | 14 | | 36.000 | 36.000 | 36.000 | |
| 6 | 31 | | | 42.613 | 42.613 | 42.613 |
| 7 | 35 | | | 46.657 | 46.657 | 46.657 |
| 8 | 50 | | | | 49.440 | 49.440 |
| 9 | 48 | | | | 51.229 | 51.229 |
| 10 Very strongly agree | 94 | | | | | 54.564 |
| Significance | | 0.056 | 0.053 | 0.080 | 0.079 | 0.262 |

Consistent with expectation, the analysis indicated there was a significant difference across the mean scale scores [$F(10, 290) = 18.478$, $p \ll 0.0001$] depending on their level of agreement or disagreement with the statement of the independent variable "*I can explain how eclipses occur.*" This led us to reject the null hypothesis of no difference. In probing the differences between subgroups on the IV, the Student-Newman-Keuls (SNK) multiple comparisons test was employed. In any multiple comparisons examination of the differences between groups of scores, if one was to extensively check for differences amongst the 11 response groups (0 to 10) on the IV, then there would be 55 possible paired comparisons usually computed using separate $t$ tests. However, one can no longer accept the $p$ value of 0.05 as the level below which a difference between any two subgroups on the independent variable (IV) is significant as one might do for a single pair because the confidence level that the difference is real is very small, that is, the confidence level of a difference in any of the 55 separate univariate analyses of pairs of groups falls to $(1-0.05)^{55}$ given that the IV has 11 response categories.

To maintain the $p < 0.05$ for each of those multiple comparisons, one would need to drop the $p$ level to 0.05/55 or $p < 0.0009$ in order to claim that the difference between the mean scores of any two subgroups is significant. This is described as a "full Bonferroni correction." It is notoriously conservative [122].

There are many other procedures in which the $p$ value can be protected to guard against a type 1 error, that is, saying that there is a significant difference between the mean scores of any two subgroups when in fact there is not. These multiple-comparison protections include Duncan, Scheffé, Tukey, or Dunnets T3, etc., each with statistical assumptions that the researcher must meet. These are available in packages such as SPSS. One of these, the SNK protection is not dependent on the individual group sizes and lies in the middle of the level of conservativeness from computing multiple $t$ tests and accepting the $p < 0.5$ (not at all conservative and with the many dangers of multiple opportunities for making a type 1 error) through to employing a full Bonferroni correction [very conservative (i.e., a $p < 0.05/55$)]. With SNK the $p$ level is protected but it is neither overly conservative nor too liberal. Table V thus presents the output for the multiple-comparisons analysis using a SNK protection.

The results indicate a clear pattern of reduction in the mean scores of APSE corresponding with a decrease in the strength of agreement with the above statement used as the IV. Students who most strongly agree with the statement "*I can explain how eclipses occur*" have the highest personal self-efficacy scores (means = 54.6 and 51.2 for the two highest scoring groups). The five groups that are found in this same column of Table V indicates that the group means of those who ticked 6,7, 8, 9, or 10 are not significantly different from each other. This "Group 5" is described as a "homogeneous subset" whose mean scores do not significantly differ from each other. The $p$ level for the differences is given at the bottom of the column, which in this case is $p = 0.262$. Nonetheless, the mean personal self-efficacy scores decrease as the level of agreement with the IV statement decreases.

The next homogeneous subset whose mean scores are not significantly different from each other is indicated in column 4 of the Table V and comprises those who ticked 5, 6, 7, 8, or 9 on the IV. However, the overall mean score for this homogenous subset is significantly different from the subset found in column 5 at the p-level of 0.05 shown across the top of the five columns of mean scores.

TABLE VI.  Multiple comparisons using SNK for homogeneous subgroups of ISE based on the item indicated. Note that means for groups in homogeneous subsets are displayed.

| Q22 I could identify objects that are moving across the sky by examining a series of astronomical images | N | ISE | | | | | |
|---|---|---|---|---|---|---|---|
| | | Subset for $\alpha = 0.05$ | | | | | |
| | | 1 | 2 | 3 | 4 | 5 | 6 |
| 0 | 6 | 6.000 | | | | | |
| 1 | 7 | | 16.857 | | | | |
| 3 | 10 | | 21.800 | 21.800 | | | |
| 2 | 9 | | | 26.333 | 26.333 | | |
| 5 | 29 | | | 29.621 | 29.621 | | |
| 4 | 18 | | | | 34.333 | 34.333 | |
| 6 | 29 | | | | 35.103 | 35.103 | |
| 7 | 37 | | | | | 39.973 | 39.973 |
| 8 | 41 | | | | | 42.390 | 42.390 |
| 9 | 41 | | | | | | 45.634 |
| 10 | 74 | | | | | | 48.014 |
| Significance | | 1.000 | 0.158 | 0.067 | 0.061 | 0.099 | 0.100 |

Thus the 11 subgroups who responded 0 to 10 on the IV are formed into five "homogeneous subsets" within which the mean scores are not significantly different from each other and with the significance level given at the bottom of each column for each homogenous subset. However, each of these five homogeneous subsets is significantly different from the others as indicated by the 0.05 level at the top of the five groups.

Therefore, we can conclude that the scale hypothesized to measure Astronomical Personal Self-Efficacy possesses a very high level of construct validity. It is measuring what it is hypothesized to be measuring: the level of students' APSE.

### 2. Testing the construct validity of the Astronomy Instrumental Self-Efficacy scale

When students learn how to use remote telescopes, one of the concepts inherent in the learning is the idea that almost all the objects in the sky appear to move across it in a regular fashion, while certain objects such as planets, the Moon, spacecraft and meteors, move differently relative to the background stars and galaxies. We hypothesized that students who possess a high sense of self-efficacy around the use of remote telescopes would more strongly agree with the statement "I could identify objects that are moving across the sky by examining a series of astronomical images" as the IV compared with those who disagree. Again, the null hypothesis for this analysis is that there will be no difference in the mean scores across the response categories of the independent variable.

We again computed a one-way ANOVA with *post-hoc* multiple comparisons with SNK protection to test this hypothesis. Consistent with expectation, the analysis indicated there is a significant difference in the mean scale scores [$F(10, 290) = 28.495$, $p \ll 0.0001$] leading to reject the null hypothesis. Table VI, again using Student-Newman-Keuls protection, identifies the homogeneous subsets and shows that with the exceptions of subgroups who ticked 2 and 4 on the independent variable, as the level of agreement with the above statement increases so does the mean ISE score. We can conclude, therefore, that the ISE scale possesses high level of construct validity.

### 3. Testing ISE and APSE with cohort A and cohort B

As indicated in Sec. V.B.1, we extracted a random sample of 58 cases from cohort A to match the $N$ of cohort B. We created a new IV to indicate membership of each cohort.

We hypothesized that the APSE would be less likely to show any differences across the two cohorts given that both were enrolled in an astronomy course at the same level of study in the same semester at the same university. Here, the null hypothesis is that there is no significant difference in the mean scores of the APSE scale. In contrast, we hypothesized that the ISE scale score would likely show major differences given that cohort A had used robotic telescopes as part of their coursework while cohort B had not. Thus, we hypothesized that the random sample of 58 cases taken from cohort A would have a higher self-efficacy in ISE than the 58 cases in cohort B. The null hypothesis in this case is that there is no statistical difference in the mean scores of the two cohorts.

In all of the distributions shown in Figs. 3 and 4, a normal distribution overlays the bar chart. Figure 3

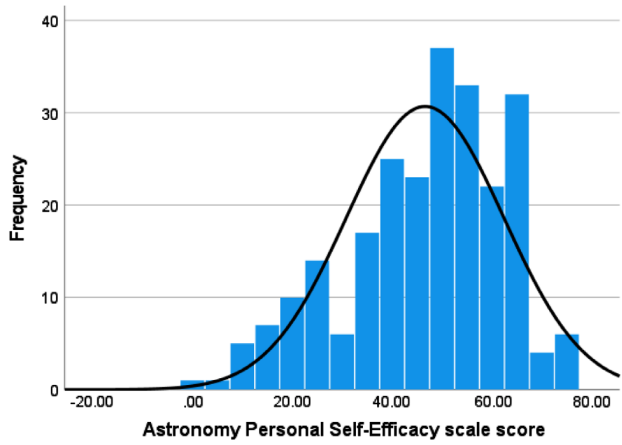

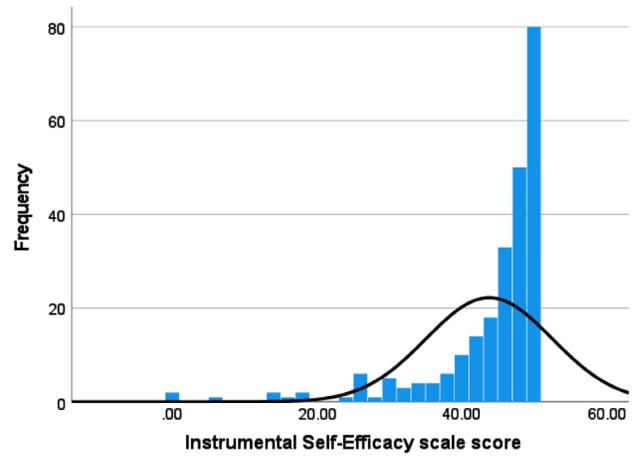

Mean = 46.60, Std. Dev. = 15.799, N = 243          Mean = 43.81, Std. Dev. = 8.724, N = 243

FIG. 3.   Lab-group frequency distribution of the APSE (left) and ISE (right) for cohort A.

illustrates the distributions of the total scores for APSE and ISE scales for cohort A. In all distributions in Figs. 3 and 4, a normal distribution overlays the bar chart. The frequency distribution of responses for the ISE scale is highly skewed and is also likely constrained by a ceiling effect, with a mean score of approximately 44 out of 50. We hypothesized further that the ceiling effect could be explained since the cohort A students had already used robotic telescopes in their lab course. The ISE construct is a more context-specific one than the APSE, which involves conceptual change about astronomical phenomena. Moreover, once a skill is learned and practiced in the lab, it may not be very easily forgotten. This behavioral change is symptomatic of real learning.

The distributions of the two scales for cohort B are shown in Fig. 4. The skewness and kurtosis of the Instrumental Self-Efficacy score for this group are much better compared with cohort A who had already used telescopes and associated instrumentation.

Our hypotheses that cohort B would have a significantly lower ISE mean score compared with the students in cohort A due to their nonuse of telescopes and instrumentation, and that there should be no significant difference in the APSE mean scores of those who had or had not used remote telescopes as this variable is based on their understanding of the same astronomy content covered in both courses were tested using two one-way ANOVAS using cohort membership as the IV.

Table VII presents the descriptive statistics for these two equal-sized groups involving the two scales. This shows that there is little difference in the APSE of the two cohorts while there appears to be a large difference in the ISE scores.

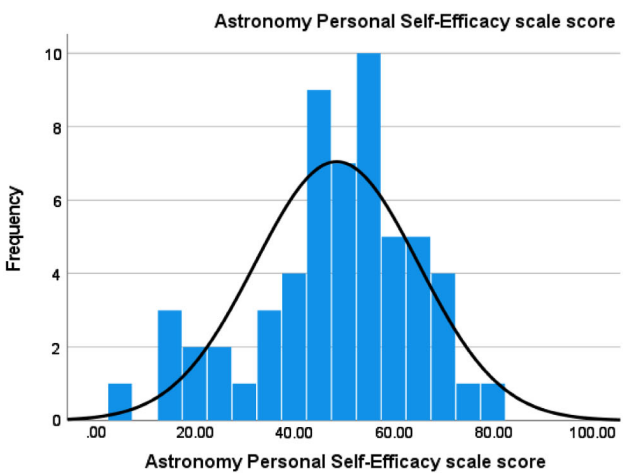

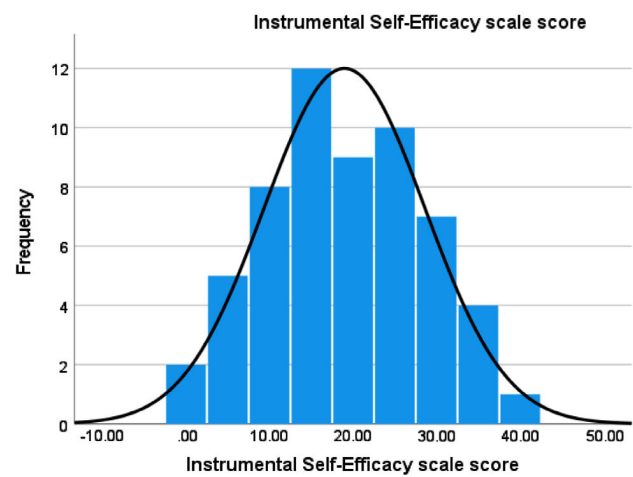

Mean = 48.57, Std. Dev. = 16.42, N = 58          Mean = 18.97, Std. Dev. = 9.641, N = 58

FIG. 4.   Nonlab group frequency distribution of the APSE (left) and ISE (right) for cohort B.

TABLE VII.   Descriptive statistics of APSE and ISE for cohort A and cohort B.

|  | N | Mean | Standard deviation | Standard error | Lower bound | Upper bound | Minimum | Maximum |
|---|---|---|---|---|---|---|---|---|
| Astronomy personal self-efficacy |  |  |  |  |  |  |  |  |
| Lab group | 58 | 46.259 | 13.380 | 1.757 | 42.740 | 49.777 | 15.00 | 69.00 |
| Nonlab group | 58 | 49.690 | 15.507 | 2.036 | 45.612 | 53.767 | 5.00 | 78.00 |
| Total | 116 | 47.974 | 14.522 | 1.348 | 45.303 | 50.645 | 5.00 | 78.00 |
| Instrumental self-efficacy |  |  |  |  |  |  |  |  |
| Lab group | 58 | 43.793 | 7.261 | 0.953 | 41.884 | 45.702 | 15.00 | 50.00 |
| Nonlab group | 58 | 20.293 | 9.599 | 1.260 | 17.769 | 22.817 | 3.00 | 41.00 |
| Total | 116 | 32.043 | 14.528 | 1.349 | 29.371 | 34.715 | 3.00 | 50.00 |

TABLE VIII.   One-way ANOVAs of the APSE and ISE between cohort A and cohort B.

|  |  | Sum of squares | df | Mean square | $F$ | Significance |
|---|---|---|---|---|---|---|
| Instrumental self-efficacy | Between groups | 16015.250 | 1 | 16015.250 | 221.100 | 0.000[*] |
|  | Within groups | 8257.534 | 114 | 72.435 |  |  |
|  | Total | 24272.784 | 115 |  |  |  |
| Astronomy personal self-efficacy | Between groups | 341.388 | 1 | 341.388 | 1.628 | 0.205 |
|  | Within groups | 23911.534 | 114 | 209.750 |  |  |
|  | Total | 24252.922 | 115 |  |  |  |

[*]Note.  $p \ll 0.0001$.

We subsequently computed the two one-way ANOVAs to determine if the observed differences in the mean scores of the two scales displayed in Table VII were statistically significant. Summaries of the outputs are in Table VIII. A full-Bonferroni protection for the $p$ value was employed because two separate univariate computations had been computed separately on the DVs. Thus, the new $p$ value below which significance can be claimed is $p < 0.025$.

Table VIII demonstrates that there is a highly significant difference in the ISE scores of cohorts A and B [$F(1, 114) = 221.100$, $p \ll 0.0001$] with a very large effect size of greater than $2\sigma$ (Cohen's $d = 2.28$). We can, with confidence, reject the null hypothesis of no difference in this case. These results suggests that the ISE scale can validly distinguish between those who have engaged in lab participation using robotic telescopes and

TABLE IX.   Final survey items.

Astronomy personal self-efficacy
1 I can do astronomy
2 I can explain how the length of the day changes with latitude
3 Most astronomy concepts are easy to learn
4 I can explain how spectroscopy works
5 I can explain why stars are different colors and brightness
6 I have a good grasp of what objects exist within and around our galaxy
7 The current scientific model of the origin and evolution of the universe is clear to me
8 Given appropriate information about standard candles (RR Lyrae, Cepheids or Type 1a Supernovae), I can calculate their distance

Instrumental Self-Efficacy
9 Adjusting the brightness and contrast levels in astronomical images is straightforward
10 I am able to request telescope images through a web-based portal
11 Selecting different filters for a remote telescope observation is easy
12 I can show someone how to request an image from a remote telescope using an online portal
13 I can learn how to use a remote telescope those who have not. In contrast, we cannot reject the null hypothesis of "no difference" in the means of the two cohorts for the APSE scale. Table VIII shows that there is no significant difference in the APSE scores showing that both groups were almost equally assured about their astronomical knowledge [$F(1, 114) = 1.628$], $p = 0.205$.

The survey items for both constructs are shown in Table IX.

## VII. DISCUSSION

The purpose of the current study was to develop an astronomy self-efficacy instrument with a focus on using robotic telescopes and to test any scales that emerged for reliability and validity. The unconstrained exploratory factor analysis yielded five factors with several of the factors initially containing only two items. Inspection of the pattern matrix and the scree plot seemed to indicate that there were only two or three factors at play. Subsequently, we computed a constrained EFA limited to two factors that yielded interpretable results. We interpreted the two constructs to be respondents' self-efficacy with respect to the content knowledge they possessed (Astronomy Personal Self-Efficacy) and their self-efficacy with respect to their use of remote telescopes (Instrumental Self-Efficacy).

During the analysis, we observed an apparent disconnect between taking an astronomy class and a feeling that there is anything related to ever becoming an astronomer. In other words, students may be thinking "I'm taking an astronomy class, but it doesn't really relate to being an astronomer." Hazari *et al.* [67] used the following example: "I am a physics person because I love learning about relativity" (p. 983) to exemplify the context to develop their physics identity. Given the desirability of developing a strong sense of science identity in this astronomy context and the relationship between it and self-efficacy, an important first step is to develop a reliable and valid way of measuring astronomy self-efficacy. Similarly, Hazari *et al.* [67] mention that physics identity is seen to be developed through a person's personal and social sense of self through their tangible experiences with physics. Therefore, we wrote three items referring to astronomers in general. However, the factor loadings on all three questions were quite low ($<0.485$) despite self-efficacy around astronomy concepts and telescope use being quite high.

The students who piloted this survey instrument were in a typical introductory astronomy course either with or without a lab component. Research has shown that science learned in a highly structured classroom environment may become strongly associated with that formal setting thereby prohibiting the learning from being applied to the wider picture [123,124]. In this case, learning in the structured environment of an introductory astronomy course may not translate to the real world of astronomy and astronomers' work and research in the minds of the students. A next step would be to apply this survey instrument more widely to students who are undertaking a variety of research experiences where there is a less "classroom-type" structured setting. In this way, we hope that these students may experience a greater connection to the way science is carried out in a real way where scientific questions can be explored.

The results of this study have shown that self-efficacy can be measured using a specialized astronomy self-efficacy instrument. The experience that participants have had during their engagement with the program involved both mastery experiences and vicarious experiences with the use of robotic telescope equipment and these two dimensions may well be responsible for the differences in self-efficacy that this instrument measured. Students who had used the remote telescopes a few times quickly gained a sense of mastery over their use. The mastering of astronomy content is expected to take longer and thus we see no significant difference in the construct of astronomy personal self-efficacy.

## VIII. CONCLUSION

Years of research into research experiences for undergraduates has shown that these enriched activities can improve a students' motivation in science, their sense of belonging to a community, their science identity and self-efficacy in specific relevant domains leading to their persistence in STEM. As we move forward in education, especially in a time when the global community is forced to find online solutions to education problems, creating research experiences for students is ever more crucial. Equally, an understanding of the components of these experiences that have the most impact is critical. Self-efficacy, while being domain specific, is well known to be an important part of the learning process and so creating programs that increase self-efficacy within science and technology courses is incumbent upon curriculum designers and those who implement research-experience programs for undergraduates. This study highlights the importance of the development of a domain-specific self-efficacy instrument as a precursor to probe further science identity as a link to future STEM career pathways. The instrument presented here reliably and validly measures two aspects of astronomy self-efficacy. Astronomical Personal Self-Efficacy is content focused and is related to students' perception of their *understanding* of astronomy and astronomy concepts they perceive themselves to have. Instrumental Self-Efficacy is based on the use of robotic telescopes and associated technologies, or the *doing* of astronomy. This tool will allow different programs to be examined and to develop highly effective pedagogical approaches to help build a more scientifically literate populace entering the workforce at a time when this is paramount.

## IX. LIMITATIONS

This instrument is limited in the sense that it does not explicitly address all four components of self-efficacy as laid out by Bandura [28]. However, it is focused on astronomy conceptual understanding and students' confidence in their beliefs through mastery experiences related to robotic telescope use. As participants work with others in teams during their programs, other factors such as vicarious experiences, social persuasion, and physiological state are implicitly embedded. Further development of the instrument could include explicit questions to address these other factors. Further research should also involve a confirmatory factor analysis to validate the factor structure suggested in this exploratory work.

The ISE scale was highly skewed in the case of those who had already used robotic telescopes [cohort A]. This is a problem that will need to be addressed in further research and perhaps in creating additional items that could produce a normally distributed scale of student responses. The purpose of this would be to employ multivariate analysis of variance procedures to probe any effects of labs that employ the use of robotic telescopes and its impact on different subgroups, for example, different genders, underrepresented minority groups, and different academic levels.

## ACKNOWLEDGMENTS

We would like to thank Professor Dan Reichart for his participation in assisting the research team to gain access to meaningful data. We gratefully acknowledge the support of the National Science Foundation, through IUSE Grant No. 2013300, and the Department of Defense, through NDEP Grant No. HQ00342110018.

[1] American Association for the Advancement of Science, *Benchmarks for Science Literacy* (AAAS, Washington, DC, 1993). Retrieved November 15, 2020, https://www.aaas.org/resources/benchmarks-science-literacy.

[2] NGSS Lead States, *Next Generation Science Standards: For States, by States* (The National Academies Press, Washington DC, 2013). Retrieved from https://www.nextgenscience.org.

[3] J. Osborne, S. Simon, and S. Collins, Attitudes towards science: A review of the literature and its implications, Int. J. Sci. Educ. **25**, 1049 (2003).

[4] National Research Council, *A Framework for K-12 Science Education: Practices, Cross-Cutting Concepts, and Core Ideas* (National Academies Press, Washington, DC, 2012).

[5] National Science Foundation, *Science, and Engineering Indicators 2018* (National Science Foundation, Washington, DC, 2018). Retrieved from February 28, 2019, https://nsf.gov/statistics/2018/nsb20181/report/sections/elementary-and-secondary-mathematics-and-science-education/highlights.

[6] K. Williamson, T. Rector, and J. Lowenthal, Embedding climate change engagement in astronomy education and research, An independent submission to Astro2020, the Astronomy and Astrophysics Decadal Survey (2019), https://arxiv.org/abs/1907.08043.

[7] T. M. Akram, A. Ijaz, and H. Ikram, Exploring the factors responsible for declining students' interest in chemistry, Int. J. Info. Educ. Technol. **7**, 88 (2017).

[8] S. Tröbst, T. Kleickmann, K. Lange-Schubert, A. Rothkopf, and K. Möller, Instruction and students' declining interest in science: An analysis of German fourth and sixth-grade classrooms, Am. Educ. Res. J. **53**, 162 (2016).

[9] J. H. Falk, N. Staus, L. D. Dierking, W. Penuel, J. Wyld, and D. Bailey, Understanding youth STEM interest pathways within a single community: The Synergies project, Int. J. Sci. Educ., Part B **6**, 369 (2016).

[10] A. Krapp and M. Prenzel, Research on interest in science: Theories, methods, and findings, Int. J. Sci. Educ. **33**, 27 (2011).

[11] A. Krapp, Basic needs and the development of interest and intrinsic motivational orientations, Learning Instr. **15**, 381 (2005).

[12] J. N. Schinske, H. Perkins, A. Snyder, and M. Wyer, Scientist spotlight homework assignments shift students' stereotypes of scientists and enhance science identity in a diverse introductory science class, CBE Life Sci. Educ. **15**, ar47 (2016).

[13] L. Martin-Hansen, Examining ways to meaningfully support students in STEM, Int. J. STEM Educ. **5**, 53 (2018).

[14] A. Y. Kim and G. M. Sinatra, Science identity development: An interactionist approach, Int. J. STEM Educ. **5**, 51 (2018).

[15] M. Beier and A. Rittmayer, *Literature overview: Motivational factors in STEM: Interest and self-concept* (Assessing Women and Men in Engineering, 2008), http://aweonline.org/arp_selfconcept_overview_122208_002.pdf.

[16] S. Glynn, P. Brickman, N. Armstrong, and G. Taasoobshirazi, Science Motivation Questionnaire II: Validation with science majors and nonscience majors, J. Res. Sci. Teach. **48**, 1159 (2011).

[17] S. Glynn, G. Taasoobshirazi, and P. Brickman, Science motivation questionnaire: Construct validation with non-science majors, J. Res. Sci. Teach. **46**, 127 (2009).

[18] P. Potvin and A. Hasni, Interest, motivation and attitude towards science and technology at K-12 levels: a systematic review of 12 years of educational research, Studies Sci. Educ. **50**, 85 (2014).

[19] L. Archer, J. DeWitt, J. Osborne, J. Dillon, B. Willis, and B. Wong, "Doing" science versus "being" a scientist: Examining 10/11-year-old schoolchildren's constructions of science through the lens of identity, Sci. Educ. 94, 617 (2010).

[20] J. Deehan, The Science Teaching Efficacy Belief Instruments (STEBI A and B): A Comprehensive Review of Methods and Findings from 25 Years of Science Education Research (Springer International Publishing, New York, 2016).

[21] M.-T. Wang and J. L. Degol, Gender gap in science, technology, engineering, and mathematics (STEM): Current knowledge, implications for practice, policy, and future directions, Educ. Psychol. Rev. 29, 119 (2017).

[22] P. Doerschuk, C. Bahrim, J. Daniel, J. Kruger, J. Mann, and C. Martin, Closing the gaps and filling the STEM pipeline: A multidisciplinary approach, J. Sci. Educ. Technol. 25, 682 (2016).

[23] M. Dembo and S. Gibson, Teachers' sense of efficacy: An important factor in school improvement, Elementary School J. 86, 173 (1985).

[24] A. L. Govett, Teachers' conceptions of the nature of science: Analyzing the impact of a teacher enhancement program in changing attitudes and perceptions of science and scientific research, Ph.D. thesis, West Virginia University, Morgantown, WV, 2001.

[25] G. Shroyer, I. Riggs, and L. Enochs, The relationship of pupil control to preservice elementary science teacher self–efficacy and outcome expectancy, Sci. Educ. 79, 63 (1995).

[26] M. Stears, A. James, and M.-A. Good, Teachers as learners: A case study of teachers' understanding of astronomy concepts and processes in an ACE course, S. Afr. J. High. Educ. 25, 568 (2011).

[27] A. Bandura, Self-efficacy: The Exercise of Control (W.H. Freeman and Company, New York, 1997).

[28] M. Feder, One Decade, One Million more STEM Graduates 25, 2019 (2012), from https://obamawhitehouse .archives.gov/blog/2012/12/18/one-decade-one-million-more-stem-graduates.

[29] A. Bandura and R. H. Walters, Social Learning Theory (Prentice- Hall Englewood Cliffs, NJ, 1977), Vol. 1.

[30] A. Bandura, Guide for constructing self-efficacy scales, Self-Efficacy Beliefs of Adolescents 5, 307–337 (2006).

[31] E. Seymour and N. M. Hewitt, Talking about Leaving: Why Undergraduates Leave the Sciences (Westview Press, Boulder, CO, 1997).

[32] A. Hasni and P. Potvin, Student's interest in science and technology and its relationships with teaching methods, family context and self-efficacy, Int. J. Environ. Sci. Educ. 10, 337 (2015).

[33] D. P. Cartrette and B. M. Melroe-Lehrman, Describing changes in undergraduate students' preconceptions of research activities, Res. Sci. Educ. 42, 1073 (2012).

[34] L. J. Rennie, D. Goodrum, and M. Hackling, Science teaching and learning in Australian schools: Results of a national study, Res. Sci. Educ. 31, 455 (2001).

[35] L. Smyth, F. Davila, T. Sloan, E. Rykers, S. Backwell, and S. B. Jones, How science really works: The student experience of research-led education, Higher Educ. 72, 191 (2016).

[36] Council on Undergraduate Research Issues Updated Definition of Undergraduate Research, | General News—News | Council on Undergraduate Research (2021), https://www .cur.org/council_on_undergraduate_research_issues_ updated_definition_of_undergraduate_research/.

[37] M. C. Linn, E. Palmer, A. Baranger, E. Gerard, and E. Stone, Undergraduate research experiences: Impacts and opportunities, Science 347, 1261757 (2015).

[38] C. B. Russell and G. C. Weaver, A comparative study of traditional, inquiry-based, and research-based laboratory curricula: Impacts on understanding of the nature of science, Chem. Educ. Res. Pract. 12, 57 (2011).

[39] O. A. Adedokun, A. B. Bessenbacher, L. C. Parker, L. L. Kirkham, and W. D. Burgess, Research skills and STEM undergraduate research students' aspirations for research careers: Mediating effects of research self-efficacy, J. Res. Sci. Teach. 50, 940 (2013).

[40] T. Perez, J. G. Cromley, and A. Kaplan, The role of identity development, values, and costs in college STEM retention, J. Educ. Psychol. 106, 315 (2014).

[41] R. D. Robnett, M. M. Chemers, and E. L. Zurbriggen, Longitudinal associations among undergraduates' research experience, self-efficacy, and identity, J. Res. Sci. Teach. 52, 847 (2015).

[42] W. K. Adams, K. K. Perkins, N. S. Podolefsky, M. Dubson, N. D. Finkelstein, and C. E. Wieman, New instrument for measuring student beliefs about physics and learning physics: The Colorado learning attitudes about science survey, Phys. Rev. ST Phys. Educ. Res. 2, 010101 (2006).

[43] M. J. Graham, J. Frederick, A. Byars-Winston, A. B. Hunter, and J. Handelsman, Increasing persistence of college students in STEM, Science 341, 1455 (2013).

[44] A. S. Trotter, D. E. Reichart, and A. P. LaCluyze, Factors contributing to attitudinal gains in introductory astronomy courses, in Proceedings of Robotic Telescopes, Student Research and Education 2 (2018).

[45] J. S. Eccles, Understanding women's educational, and occupational choices: Applying the Eccles, et al. model of achievement-related choices, Psychol. Women Q. 18, 585 (1994).

[46] R. W. Lent, S. D. Brown, and G. Hackett, Toward a unifying social cognitive theory of career and academic interest, choice, and performance, J. Vocat. Behav. 45, 79 (1994).

[47] J. R. Cordova, G. M. Sinatra, S. H. Jones, G. Taasoobshirazi, and D. Lombardi, Confidence in prior knowledge, self-efficacy, interest and prior knowledge: Influences on conceptual change, Contemp. Educ. Psychol. 39, 164 (2014).

[48] J. A. Dole and G. M. Sinatra, Reconceptualizing change in the cognitive construction of knowledge, Educational Psychol. 33, 109 (1998).

[49] S. Bartlett, M. T. Fitzgerald, D. H. McKinnon, L. Danaia, and J. Lazendic-Galloway, Astronomy and Science Student Attitudes (ASSA): A short review and validation of a new instrument, J. Astron. Earth Sci. Educ. 5, 1 (2018).

[50] A. Bandura, On the functional properties of perceived self-efficacy revisited, J. Manage. 38, 9 (2012).

[51] V. Sawtelle, E. Brewe, and L. H. Kramer, Exploring the relationship between self-efficacy and retention in introductory physics, J. Res. Sci. Teach. 49, 1096 (2012).

[52] R. E. Bleicher, Revisiting the STEBI-B: Measuring self-efficacy in preservice elementary teachers, School Sci. Math. 104, 383 (2004).

[53] M. Tschannen-Moran, A. W. Hoy, and W. K. Hoy, Teacher efficacy: Its meaning and measure, Rev. Educ. Res. 68, 202 (1998).

[54] R. E. Bleicher, Nurturing confidence in preservice elementary science teachers, J. Sci. Teach. Educ. 18, 841 (2007).

[55] N. G. Lederman and J. S. Lederman, The status of preservice science teacher education: A global perspective, J. Sci. Teach. Educ. 26, 1 (2015).

[56] C. M. Norris, Exploring the impact of postgraduate preservice primary science education on students' self-efficacy, Ph.D. thesis, Edith Cowan University, 2017 (unpublished).

[57] M. Braund and M. Driver, Pupils' perceptions of practical science in primary and secondary school: Implications for improving progression and continuity of learning, Educ. Res. 47, 77 (2005).

[58] A. Emvalotis and A. Koutsianou, Greek primary school students' images of scientists and their work: Has anything changed?, Res. Sci. Technol. Educ. 36, 69 (2018).

[59] C. P. Tan, H. t. Van der Molen, and H. g. Schmidt, A measure of professional identity development for professional education, Studies Higher Educ. 42, 1504 (2017).

[60] J. Zhai, J. A. Jocz, and A.-L. Tan, 'Am I Like a Scientist?': Primary children's images of doing science in school, Int. J. Sci. Educ. 36, 553 (2014).

[61] J. E. Stets and P. J. Burke, Identity theory and social identity theory, Social Psychol. Quart. 63, 224 (2000).

[62] B. A. Brown, J. M. Reveles, and G. J. Kelly, Scientific literacy and discursive identity: A theoretical framework for understanding science learning, Sci. Educ. 89, 779 (2005).

[63] A. Wigfield and J. S. Eccles, Expectancy-value theory of achievement motivation, Contemp. Educ. Psychol. 25, 68 (2000).

[64] J. Eccles, Special Issue, Who am I and what am I going to do with my life? Personal and collective identities as motivators of action, Motivation and Identity 44, 78 (2009).

[65] A. V. Maltese and R. H. Tai, Eyeballs in the fridge: Sources of early interest in science, Int. J. Sci. Educ. 32, 669 (2010).

[66] A. Colantonio, I. Marzoli, E. Puddu, S. Bardelli, M. T. Fulco, S. Galano, L. Terranegra, and I. Testa, Describing astronomy identity of upper primary and middle school students through structural equation modeling, Phys. Rev. Phys. Educ. Res. 17, 010139 (2021).

[67] Z. Hazari, G. Sonnert, P. M. Sadler, and M.-C. Shanahan, Connecting high school physics experiences, outcome expectations, physics identity, and physics career choice: A gender study, J. Res. Sci. Teach. 47, 978 (2010).

[68] D. Lopatto, Undergraduate research experiences support science career decisions and active learning, CBE Life Sci. Educ. 6, 297 (2007).

[69] E. Wenger, Communities of Practice: Learning, Meaning, and Identity (Cambridge University Press, Cambridge, England, 1999).

[70] D. I. Hanauer and E. L. Dolan, The Project ownership survey: Measuring differences in scientific inquiry experiences, CBE Life Sci. Educ. 13, 149 (2014).

[71] D. K. C. Trundle, Teaching science during the early childhood years, 5. National Geographic Learning, Cengage. Boston, MA, http://ngspscience.com/profdev/Monographs/SCL22-0429A_SCI_AM_Trundle_lores.pdf.

[72] K. Green, K. C. Trundle, and M. Shaheen, Integrating the arts into science teaching and learning: A literature review, J. Learning through the Arts 14 (2019).

[73] N. A. Falk, P. J. Rottinghaus, T. N. Casanova, F. H. Borgen, and N. E. Betz, Expanding women's participation in STEM: Insights from parallel measures of self-efficacy and interests, J. Career Assess. 25, 571 (2017).

[74] L. T. Ko, R. R. Kachchaf, A. K. Hodari, and M. Ong, Agency of women of color in physics and astronomy: strategies for persistence and success, J. Women Minorities Sci. Engineering 20, 171 (2014).

[75] J. C. Palmer, The efficacy of planetarium experiences to teach specific science concepts, Ph.D. thesis, Texas A&M University, College Station, TX, 2007.

[76] M. Porche, J. M. Grossman, and K. C. Dupaya, New American scientists: First generation immigration status and college STEM aspirations, J. Women Minorities Sci. Engineering 22, 1 (2016).

[77] C. van Tuijl and J. H. W. van der Molen, Study choice and career development in STEM fields: An overview and integration of the research, Int. J. Technol. Des. Educ. 26, 159 (2016).

[78] J. Legewie and T. A. DiPrete, The high school environment and the gender gap in science and engineering, Sociol. Educ. 87, 259 (2014).

[79] S. González-Pérez, R. Mateos de Cabo, and M. Sáinz, Girls in STEM: Is it a female role-model thing?, Front. Psychol. 11, 1 (2020).

[80] L. J. Kewley, Closing the gender gap in the Australian astronomy workforce, Nat. Astron. 5, 615 (2021).

[81] L. Archer, J. DeWitt, J. Osborne, J. Dillon, B. Willis, and B. Wong, Science aspirations, capital, and family habitus: How families shape children's engagement and identification with science, Am. Educ. Res. J. 49, 881 (2012).

[82] M. Ghee, M. Keels, D. Collins, C. Neal-Spence, and E. Baker, Fine-tuning summer research programs to promote underrepresented students' persistence in the STEM pathway, CBE Life Sci. Educ. 15, ar28 (2016).

[83] A. J. Gonsalves, Exploring how gender figures the identity trajectories of two doctoral students in observational astrophysics, Phys. Rev. Phys. Educ. Res. 14, 010146 (2018).

[84] R. Ivie, S. White, and R. Y. Chu, Women's and men's career choices in astronomy and astrophysics, Phys. Rev. Phys. Educ. Res. 12, 020109 (2016).

[85] A. M. Porter and R. Ivie, Women in Physics and Astronomy, 2019 Report (American Institute of Physics,

Statistics Research Center, College Park, MD, 2019), p. 45.

[86] R. S. Barthelemy, M. McCormick, and C. Henderson, Gender discrimination in physics and astronomy: Graduate student experiences of sexism and gender microaggressions, Phys. Rev. Phys. Educ. Res. **12**, 020119 (2016).

[87] F. Demirci and C. Ozyurek, Astronomy teaching self-efficacy belief scale: The validity and reliability study, J. Educ. Learning **7**, 258 (2017).

[88] J. M. Bailey, D. Lombardi, J. R. Cordova, and G. M. Sinatra, Meeting students halfway: Increasing self-efficacy and promoting knowledge change in astronomy, Phys. Rev. Phys. Educ. Res. **13**, 020140 (2017).

[89] J. S. Krim, L. E. Coté, R. S. Schwartz, E. M. Stone, J. J. Cleeves, K. J. Barry, W. Burgess, S. R. Buxner, J. M. Gerton, L. Horvath, J. M. Keller, S. C. Lee, S. M. Locke, and B. M. Rebar, Models and impacts of science research experiences: A review of the literature of CUREs, UREs, and TREs, CBE Life Sci. Educ. **18**, ar65 (2019).

[90] L. M. Rebull, T. Roberts, W. Laurence, M. T. Fitzgerald, D. A. French, V. Gorjian, and G. K. Squires, Motivations of educators for participating in an authentic astronomy research experience professional development program, Phys. Rev. Phys. Educ. Res. **14**, 010148 (2018).

[91] M. M. Wooten, K. Coble, A. W. Puckett, and T. Rector, Investigating introductory astronomy students' perceived impacts from participation in course-based undergraduate research experiences, Phys. Rev. Phys. Educ. Res. **14**, 010151 (2018).

[92] L. A. Corwin, M. J. Graham, and E. L. Dolan, Modeling course-based undergraduate research experiences: An agenda for future research and evaluation, CBE—Life Sciences Education **14**, 1 (2015).

[93] L. M. Rebull, M. T. Fitzgerald, T. Roberts, D. A. French, W. Laurence, V. Gorjian, and G. K. Squires, The NASA/IPAC Teacher Archive Research Program (NITARP), in *Proceedings of the RTSRE Conference, San Diego, CA* (2018).

[94] L. M. Rebull, D. A. French, W. Laurence, T. Roberts, M. T. Fitzgerald, V. Gorjian, and G. K. Squires, Major outcomes of an authentic astronomy research experience professional development program: An analysis of 8 years of data from a teacher research program, in *Proceedings of the RTSRE Conference, San Diego, CA* (2018).

[95] E. Gomez and M. T. Fitzgerald, Robotic telescopes in education, Astronomical Rev. **13**, 28 (2017).

[96] M. T. Fitzgerald, R. Hollow, L. M. Rebull, L. Danaia, and D. H. McKinnon, A review of high school level astronomy student research projects over the last two decades, Pub. Astron. Soc. Aust. 31, E037 (2014).

[97] R. Freed, M. Fitzgerald, R. Genet, and B. Davidson, An overview of ten years of student research and JDSO publications, in *Proceedings of the Society for Astronomical Sciences 36th Annual Symposium on Telescope Science Proceedings* (2017) pp 131–136.

[98] R. Genet *et al.*, Student scientific research within communities-of-practice, in *Proceedings of the Society for Astronomical Sciences 36th Annual Symposium on Telescope Science* (2017), pp 143–149.

[99] M. Fitzgerald, D. H. McKinnon, and L. Danaia, Inquiry-based educational design for large-scale high school astronomy projects using real telescopes, J. Sci. Educ. Technol. **24**, 747 (2015).

[100] M. Fitzgerald, D. H. McKinnon, L. Danaia, and J. Deehan, A large-scale inquiry-based astronomy intervention project: impact on students' content knowledge performance and views of their high school science classroom, Res. Sci. Educ. **46**, 901 (2016).

[101] M. Zeilik, C. Schau, N. Mattern, S. Hall, K. W. Teague, and W. Bisard, Conceptual astronomy: A novel model for teaching postsecondary science courses, Am. J. Phys. **65**, 987 (1997).

[102] A. Kareva, S. Miller, A. Foster, and C. R. James, *A Comparison of Astronomy/Science Attitudes Among Students and Secondary Teachers* (American Astronomical Society, AAS Meeting, 2014).

[103] S. J. Slater, T. F. Slater, and A. Shaner, Impact of backwards faded scaffolding in an astronomy course for pre-service elementary teachers based on inquiry, J. Geosci. Educ. **56**, 408 (2008).

[104] E. Gomez, Las Cumbres Observatory: Networking People and Telescopes for Science and Education, RTSRE Proc., 1 (2018).

[105] R. Freed, Astronomy Research Seminar Expansion and Building a Community-of-Practice, RTSRE Conference Proceedings, Hilo, HI, 2 (2019).

[106] S. Salimpour, S. Bartlett, M. T. Fitzgerald, D. H. McKinnon, K. R. Cutts, C. R. James, S. Miller, L. Danaia, R. P. Hollow, S. Cabezon, M. Faye, A. Tomita, C. Max, M. de Korte, C. Baudouin, D. Birkenbauma, M. Kallery, S. Anjos, Q. Wu, and A. Ortiz-Gil, The gateway science: A review of astronomy in the OECD school curricula, including China and South Africa, Res. Sci. Educ. **51**, 975 (2020).

[107] B. Partridge and G. Greenstein, Goals for 'Astro 101': Report on workshops for department leaders, Astron. Educ. Rev. **2**, 46 (2003).

[108] H. Fencl and K. Scheel, Engaging students: An examination of the effects of teaching strategies on self-efficacy and course climate in a nonmajors physics course, J. Coll. Sci. Teach. **35**, 20 (2005).

[109] R. W. Lent, S. D. Brown, and K. C. Larkin, Self-efficacy in the prediction of academic performance and perceived career options, J. Counsel. Psychol. **33**, 265 (1986).

[110] S. L. Britner and F. Pajares, Sources of science self-efficacy beliefs of middle school students, J. Res. Sci. Teach. **43**, 485 (2006).

[111] I. M. Riggs and L. G. Enochs, Toward the development of an elementary teacher's science teaching efficacy belief instrument, Sci. Educ. **74**, 625 (1990).

[112] R. E. Bleicher, Revisiting the STEBI-B: Measuring self-efficacy in preservice elementary teachers, School Sci. Math. **104**, 383 (2004).

[113] R. Freed, The Astronomy Research Seminar: The wide-ranging impact on student education and careers, preliminary results, in *Proceedings of the Society for Astronomical Sciences 37th Annual Symposium on Telescope Science* (2018), pp. 161–173.

[114] T. F. Slater, To telescope or not to telescope?, RTSRE Conference Proceedings, San Diego, CA, **1** (2018).

[115] R. Freed, Evaluation of the Astronomy Research Seminar, RTSRE Conference Proceedings, Hilo, HI, **2** (2019).

[116] J. L. Arbuckle, *IBM SPSS Amos 26 User's Guide*, Computer software and manual (IBM, New York, NY, 2019).

[117] R. B. Cattell, The scree test for the number of factors, Multivariate Behav. Res. **1**, 245 (1966).

[118] L. R. Fabrigar, D. T. Wegener, R. C. MacCallum, and E. J. Strahan, Evaluating the use of exploratory factor analysis in psychological research, Psychol. Methods **4**, 272 (1999).

[119] H. Taherdoost, S. Sahibuddin, and N. Jalaliyoon, *Exploratory factor analysis; concepts and theory*, Advances in Applied Mathematics, WSEAS, Gdansk, Poland, 8 (2014).

[120] H. Karami, Exploratory factor analysis as a construct validation tool: (Mis)applications in applied linguistics research, TESOL Journal **6**, 476 (2015).

[121] H. Abdi and L. Williams, Newman-Keuls test and Tukey test, *Encyclopedia of Research Design* (Sage Publications, Thousand Oaks, CA, 2010).

[122] R. J. F. M. J. Wilson, *Statistical Methods*, 3rd ed., edited by Rudolf J. Freund (Academic Press, New York, 1656).

[123] P. Cobb, Where is the mind? Constructivist and sociocultural perspectives on mathematical development, Educ. Res. **23**, 13 (1994).

[124] J. G. Greeno, J. L. Moore, and D. R. Smith, Transfer of situated learning, in *Transfer on Trial: Intelligence, Cognition, and Instruction* (Ablex Publishing, Norwood, NJ, 1993), pp. 99–167.